\documentclass{article}
\usepackage{dcase2024,amsmath,graphicx,url,times,booktabs, tabularx}
\usepackage{multirow}
\usepackage[dvipsnames]{xcolor}

\title{Representational learning for an anomalous sound detection system with source separation model}

\name{Seunghyeon Shin$^{1}$, 
      Seokjin Lee$^{1,2}$,
      }

\address{$^1$ School of Electronic and Electrical Engineering, Kyungpook National University, \\Daegu, Republic of Korea, \{sh.shin, sjlee6\}@knu.ac.kr\\          
	    $^2$ School of Electronics Engineering, Kyungpook National University, Daegu, Republic of Korea \\
 }

\begin{document}

\ninept
\maketitle

\begin{sloppy}

\begin{abstract}

The detection of anomalous sounds in machinery operation presents a significant challenge due to the difficulty in generalizing anomalous acoustic patterns. This task is typically approached as an unsupervised learning or novelty detection problem, given the complexities associated with the acquisition of comprehensive anomalous acoustic data. Conventional methodologies for training anomalous sound detection systems primarily employ auto-encoder architectures or representational learning with auxiliary tasks. However, both approaches have inherent limitations. Auto-encoder structures are constrained to utilizing only the target machine's operational sounds, while training with auxiliary tasks, although capable of incorporating diverse acoustic inputs, may yield representations that lack correlation with the characteristic acoustic signatures of anomalous conditions. We propose a training method based on the source separation model (CMGAN\cite{cmgan}) that aims to isolate non-target machine sounds from a mixture of target and non-target class acoustic signals. This approach enables the effective utilization of diverse machine sounds and facilitates the training of complex neural network architectures with limited sample sizes. Our experimental results demonstrate that the proposed method yields better performance compared to both conventional auto-encoder training approaches and source separation techniques that focus on isolating target machine signals. Moreover, our experimental results demonstrate that the proposed method exhibits the potential for enhanced representation learning as the quantity of non-target data increases, even while maintaining a constant volume of target class data.

\end{abstract}


\section{Introduction}
\label{sec:intro}

Anomalous sound detection aims to determine whether an acquired acoustic signal originates from normal or anomalous operating conditions. The diverse nature of anomalous sounds, which vary depending on the target system, necessitates separate datasets for each target system. Furthermore, the difficulty in acquiring comprehensive anomalous sound samples, coupled with the fact that acquired samples may not represent all possible anomalous states, necessitates approaching this problem as an unsupervised learning or novelty detection task.

In response to these challenges, DCASE 2024 Task 2 \cite{dc24_task} requires the development of an anomalous sound detection model capable of monitoring machine conditions without access to anomalous samples. The task provides 1,000 sound clips per machine type for model training. Additionally, to ensure generalization across varying operating environments and machine types, the system must be robust to environmental changes and operate without reliance on additional attribute information, such as operating speed or machine identification. 

Conventional approaches typically employ neural networks to extract representations from acoustic signals. These methods can be broadly categorized into two main groups. The first utilizes auto-encoder structures \cite{ae_net}, training neural networks to encode and decode input signals. Anomaly detection is then performed either by quantifying the discrepancy between the decoded output and the input, or by using the encoder output as an embedding feature for subsequent anomaly score calculation. The second category involves training neural networks on auxiliary tasks, such as classifying additional data attributes or classes \cite{clf_net}, or directly learning representations through contrastive learning techniques \cite{cont_net}.

Although source separation neural networks have been used in anomalous sound detection \cite{sep_net}, they have mainly served as preprocessing stages for other neural networks or have been trained using attribute information. To address the constraints of training without attribute information while effectively utilizing non-target class signals, we propose a novel source separation-based representational learning method.

Our experimental results demonstrate the efficacy of the proposed method. We conducted comparative tests against conventional source separation methods that estimate target class signals and auto-encoder structures that utilize only the target signal as input. The proposed method achieved a better representation when evaluated by using anomalous sound detection with the Mahalanobis distance. Moreover, we observed that the performance of our proposed method improves with increased non-target data, even when the quantity of target data remains constant.

\section{Methodology}
\label{sec:method}

\subsection{Training strategy and anomaly score calculation}

We obtain an embedding feature vector from a source separation model. The purpose of the neural network is to extract characteristics that can distinguish whether a machine's condition is normal or abnormal. We assume that normal and abnormal conditions can be distinguished from acoustic data and that would also be distinguishable in the representation of neural network. We utilized the CMGAN\cite{cmgan} neural network structure, which is an encoder-decoder structure with conformer blocks\cite{gulati2020conformer}.

In contrast to a typical auto-encoder structure, where the neural network is trained to reconstruct the desired input signal at the output, our training objective is to remove the target machine signal from the input and separate the signals from other machines. In our representation training process, the input of the neural network $X_{t,f}$ is expressed as follows:

\begin{equation}X_{t,f} = \mathcal{F}(d_{c}(t)+s\times n_{\Bar{c}}(t)),\label{prob}\end{equation}
where $d_{c}(t)$ represents the signal of the target machine class $c$ signal in time series, $n_{\Bar{c}}(t)$ represents the signal from another class $\Bar{c}$ that is not the target machine class, $s$ is the scaling factor to match the intensity of the target and other machine class signals in dB scale, and the neural network input $X_{t,f}$ in the time-frequency domain is obtained by applying the short-time Fourier transform operator $\mathcal{F}$ to the weighted sum of $d_{c}(t)$ and $n_{\Bar{c}}(t)$. We introduce a scaling factor $s$ to modulate the training objective's difficulty. Small value of $s$ make the intensity of non-target signal smaller, the source separation problem becomes more difficult, and vice versa. For the proposed system to work well, $s$ value needs to be sufficiently small, but a large $s$ value may be needed when the number of training samples or model complexity is limited.
\\
The neural network utilizes the real, imaginary, and magnitude components of the spectrogram as input.
Since we intend for our feature extractor to remove the target machine signal, we trained it to minimize the difference between the neural network's estimation output and the other class signal. We configured the training loss function $\mathcal{L}$ of the neural network as follows:



\begin{align} 
\begin{split}
\mathcal{L} = &\alpha\{\frac{1}{l}\sum\limits_{t=1}^{l}(|s\times n_{\Bar{c}}(t)-y(t)|)\}+\beta\{\frac{1}{mn}\sum\limits_{t=1}^m \sum\limits_{f=1}^{n}(|s\times N_{t,f}^R|-
\\
&{|Y_{t,f}^R}|)^2\}+\gamma\{\frac{1}{mn}\sum\limits_{t=1}^m \sum\limits_{f=1}^{n} (|s\times N_{t,f}^I|-{|Y}_{t,f}^I|)^2\},
\end{split}
\end{align}

where $y_{t}$ and $Y_{t,f}^R$, $Y_{t,f}^I$ are the output of the neural network decoder in the time series and the real and imaginary components of the time-frequency domain, respectively. $N_{t,f}^R$ and $N_{t,f}^I$ represent the real and imaginary components of the non-target signal in the time-frequency domain. $\alpha$, $\beta$, and $\gamma$ are the hyperparameters for each difference term. During the training procedure, the network is trained to estimate $n_{\Bar{c}}(t)$ from $X_{t,f}$. 

Anomaly scores are calculated from the Mahalanobis distance of the neural network output feature matrix. The covariance matrix of the feature is estimated by the maximum-likelihood covariance estimator.
In summary, we first train the neural network to separate the signal from the mixed signal of other classes and the target class signal, excluding the target class signal. After training, the average pooling is performed in multiple stages of the network, and the average pool is executed from the output of the encoder, the intermediate of the conformer, and the output of the conformer. The resulting average-pooled matrices are then used as features for anomaly scoring, and in the scoring process, we employ the Mahalanobis distance with a covariance estimator. The overview of our system is shown in Fig. \ref{fig:overall}.

\begin{figure}
    \centering
    \includegraphics[width=1.0\linewidth]{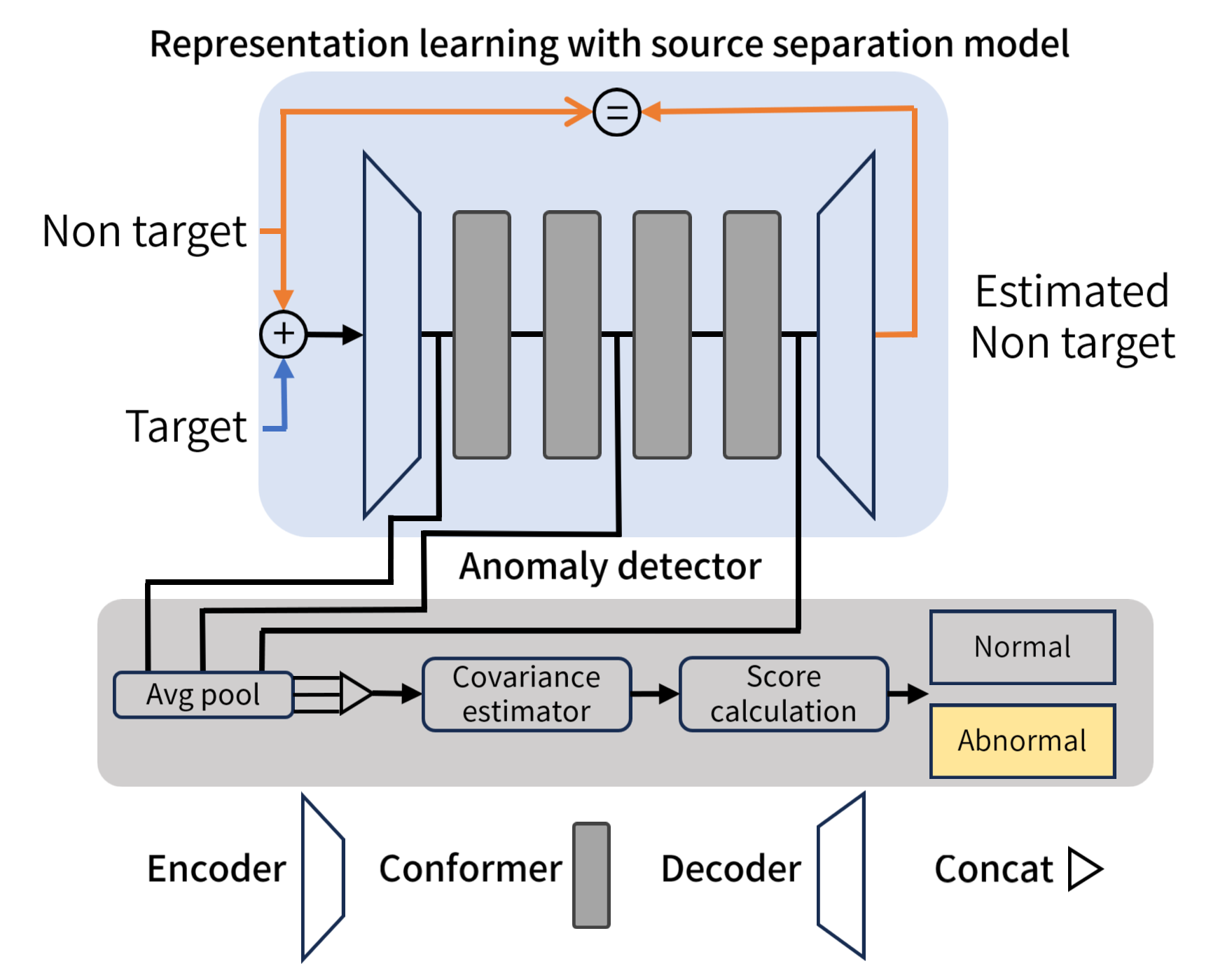}
    \caption{Proposed anomalous detection system overview}
    \label{fig:overall}
\end{figure}

\subsection{Experiments configure}

We utilized two datasets consistent with those employed in DCASE 2024 Task 2: ToyADMOS2 \cite{Harada2021} and MIMII DG \cite{Dohi2022}. These datasets collectively provided 16 types of machine sounds, each comprising 1,000 training clips. Each set of 1,000 clips was composed of 990 clips from the source domain and 10 clips from the target domain which are different operating environment from the source domain.
For evaluation purposes, 7 machine type signals were provided, consisting of 200 sound clips labeled as either normal or anomalous. The evaluation data were equally distributed between the source and target domains, with 100 clips from each domain. The source domain, constituting 99\% of the training data, represents the primary training environment. The target domain, comprising the remaining 1\% of training data, consists of domain-shifted signals that reflect changed machine operating conditions relative to the source domain.
\\
To assess the efficacy of our proposed method, we conducted experiments using varying quantities of non-target data. Two distinct datasets were employed: the first included non-target sounds from six machine types that contained test data, while the second utilized sounds from 14 machine types, excluding the Brushless Motor class due to the presence of clipping in some samples.
Audio preprocessing involved randomly trimming each 16kHz sampled clip to 2 seconds, followed by a short-time Fourier transform using a filter length of 400 samples and an overlap of 100 samples. In the decibel matching process, non-target class signals were attenuated by 5dB lower relative to the target class signal.
We added average pooling layers to the CMGAN neural network structure, applying pooling to each channel. Given the network's 64-channel architecture, the resulting feature vector maintained a 64-dimensional size post-pooling. The average pooling was performed at three locations: the encoder output, the second conformer block output, and the last conformer block output (decoder input). The 2-second average-pooled results were concatenated to form the final feature set.
The loss function hyperparameters $\alpha$, $\beta$, and $\gamma$ were set to 0.5, 6.0, and 1.0, respectively, and remained constant across all machine classes. We initially adopted hyperparameters value from the original hyperparameters of CMGAN and subsequently fine-tuned them through experimentation. The hyperparameter values are dependent on the non-target signal intensity matching level and necessitating adjustment as matching dB changes. For optimization, we used the AdamW \cite{adamw} algorithm in conjunction with a StepLR learning rate scheduler. To ensure fair comparison, hyperparameters and neural network configurations were maintained consistently across different training methodologies.

\section{Evaluation}
\label{sec:evaluation}

\subsection{Evaluation metric}

The performance of the anomalous detection system is evaluated using the Area Under the Curve (AUC) of the Receiver Operating Characteristic (ROC) curve. AUC scores are calculated in three distinct contexts: the source domain, which generates most of the training data; the target domain, representing a domain-shifted environment; and the partial AUC (pAUC), which constrains the maximum false positive rate to 10\%, thus addressing the frequency of false alarms in practical applications. The overall model performance, consistent with the DCASE 2024 Task 2 official score $\Omega$, is computed as the harmonic mean of results across all classes, domains, and the pAUC, as follows:
\begin{equation} \Omega =  h\{AUC_{c, d}, pAUC_{c} | c\in \mathcal{C}, d\in\{source, target\} \}
,\label{prob}\end{equation}
where $h$ is the harmonic mean operator and $\mathcal{C}$ is the set of the machine types. 

\subsection{Evaluation result}

We evaluated our proposed method against two conventional approaches: a source separation method utilizing the CMGAN structure and an auto-encoder method also based on CMGAN. For comprehensive performance comparison, we have included the results of the DCASE 2024 Task 2 baseline system \cite{Harada2023firstshot} in Table \ref{tab:score}.
Table \ref{tab:score} presents the performance of various systems. Our proposed method is evaluated with two configurations: estimating 14 and 6 non-target classes from a mixture of target and non-target signals. The conventional separation approach trains a neural network to estimate the target class from a mixed signal, while the auto-encoder method reconstructs the target signal from target-only input. The DCASE 2024 Task 2 baseline systems score anomalies using the mean square error (Baseline-MSE) and the Mahalanobis distance (Baseline-Mahalanobis) between input and output.

The score $\Omega$, which represents the harmonic mean of the AUC and pAUC performance metrics, demonstrates the efficacy of our approach. The neural network trained to separate non-target signals achieved a $\Omega$ score of 54.58\% This performance surpasses both the conventional separation method, which estimates target signals and achieved 53.99\%, and the auto-encoder method, which yielded 51.41\%.
Furthermore, we observed that increasing the quantity of non-target class data enhanced the potential for acquiring better representations. Specifically, expanding the diversity of non-target classes led to an improvement in the $\Omega$ score from 54.58\% to 56.00\% when using our proposed training method.
To visualize the effectiveness of our approach, we employed the t-Distributed Stochastic Neighbor Embedding (tSNE) projection to represent the learned features of the toytrain class, as illustrated in Fig. \ref{fig:tsne}. The visualization demonstrates that our proposed method, which focuses on separating non-target class signals, yields more distinct separations between normal and anomalous samples compared to alternative methods. These alternatives, which include approaches trained to separate target class signals or traditional auto-encoder methods, show less clear differentiation in their projected representations.

\begin{table*}
\centering
\begin{tabular}{|c|c|ccccccc|c|}
\hline
\textbf{System info} &
  \textbf{Metric} &
  \textbf{bearing} &
  \textbf{fan} &
  \textbf{gearbox} &
  \textbf{slider} &
  \textbf{toycar} &
  \textbf{toytrain} &
  \textbf{valve} &
  \textbf{\begin{tabular}[c]{@{}c@{}}$\Omega$ score\\ (h-mean)\end{tabular}} \\ \hline
\multirow{3}{*}{\textbf{\begin{tabular}[c]{@{}c@{}}Proposed\\ method\\ (14 class)\end{tabular}}} &
  \textbf{AUC(Target)} &
  69.24\% &
  62.76\% &
  59.84\% &
  55.88\% &
  45.96\% &
  65.64\% &
  48.76\% &
  \multirow{3}{*}{56.00\%} \\ \cline{2-2}
 &
  \textbf{AUC(Source)} &
  61.04\% &
  58.60\% &
  68.68\% &
  65.88\% &
  44.72\% &
  76.76\% &
  46.36\% &
   \\ \cline{2-2}
 &
  \textbf{pAUC} &
  58.05\% &
  54.16\% &
  55.05\% &
  51.58\% &
  48.89\% &
  54.79\% &
  48.89\% &
   \\ \hline
\multirow{3}{*}{\textbf{\begin{tabular}[c]{@{}c@{}}Proposed\\ method\\ (6 class)\end{tabular}}} &
  \textbf{AUC(Target)} &
  69.52\% &
  63.28\% &
  62.12\% &
  47.16\% &
  44.32\% &
  65.36\% &
  47.48\% &
  \multirow{3}{*}{54.58\%} \\ \cline{2-2}
 &
  \textbf{AUC(Source)} &
  61.04\% &
  55.64\% &
  61.84\% &
  60.28\% &
  47.68\% &
  77.52\% &
  48.08\% &
   \\ \cline{2-2}
 &
  \textbf{pAUC} &
  54.16\% &
  51.42\% &
  52.84\% &
  51.58\% &
  48.68\% &
  51.84\% &
  48.79\% &
   \\ \hline
\multirow{3}{*}{\textbf{\begin{tabular}[c]{@{}c@{}}Conventional\\ separation\end{tabular}}} &
  \textbf{AUC(Target)} &
  67.96\% &
  57.00\% &
  56.36\% &
  53.72\% &
  47.00\% &
  65.08\% &
  48.00\% &
  \multirow{3}{*}{53.99\%} \\ \cline{2-2}
 &
  \textbf{AUC(Source)} &
  64.52\% &
  57.96\% &
  61.80\% &
  61.64\% &
  42.48\% &
  74.64\% &
  42.24\% &
   \\ \cline{2-2}
 &
  \textbf{pAUC} &
  56.32\% &
  48.74\% &
  51.79\% &
  51.63\% &
  47.84\% &
  53.53\% &
  48.63\% &
   \\ \hline
\multirow{3}{*}{\textbf{Auto-encoder}} &
  \textbf{AUC(Target)} &
  67.84\% &
  55.48\% &
  49.84\% &
  42.16\% &
  49.72\% &
  63.04\% &
  47.56\% &
  \multirow{3}{*}{51.41\%} \\ \cline{2-2}
 &
  \textbf{AUC(Source)} &
  57.92\% &
  57.92\% &
  53.12\% &
  44.92\% &
  43.24\% &
  76.76\% &
  39.48\% &
   \\ \cline{2-2}
 &
  \textbf{pAUC} &
  51.89\% &
  51.42\% &
  \multicolumn{1}{l}{51.05\%} &
  51.05\% &
  48.11\% &
  53.00\% &
  49.32\% &
   \\ \hline
\multirow{3}{*}{\textbf{\begin{tabular}[c]{@{}c@{}}Baseline\\ (MSE)\end{tabular}}} &
  \textbf{AUC(Target)} &
  61.40\% &
  55.24\% &
  69.34\% &
  56.01\% &
  33.75\% &
  46.92\% &
  46.25\% &
  \multirow{3}{*}{55.35\%} \\ \cline{2-2}
 &
  \textbf{AUC(Source)} &
  62.01\% &
  67.71\% &
  70.40\% &
  66.51\% &
  66.98\% &
  76.63\% &
  51.07\% &
   \\ \cline{2-2}
 &
  \textbf{pAUC} &
  57.58\% &
  57.53\% &
  55.65\% &
  51.77\% &
  48.77\% &
  47.95\% &
  52.42\% &
   \\ \hline
\multirow{3}{*}{\textbf{\begin{tabular}[c]{@{}c@{}}Baseline\\ (Mahalanobis)\end{tabular}}} &
  \textbf{AUC(Target)} &
  51.58\% &
  42.70\% &
  74.35\% &
  68.11\% &
  37.35\% &
  39.99\% &
  53.61\% &
  \multirow{3}{*}{55.02\%} \\ \cline{2-2}
 &
  \textbf{AUC(Source)} &
  54.43\% &
  79.37\% &
  81.82\% &
  75.35\% &
  63.01\% &
  61.99\% &
  55.69\% &
   \\ \cline{2-2}
 &
  \textbf{pAUC} &
  58.82\% &
  53.44\% &
  55.74\% &
  49.05\% &
  51.04\% &
  48.21\% &
  51.26\% &
   \\ \hline
\end{tabular}
\caption{Anomalous detection performance comparison of proposed method and others}
\label{tab:score}
\end{table*}

\begin{figure*}
    \centering
    \includegraphics[height=0.73\linewidth]{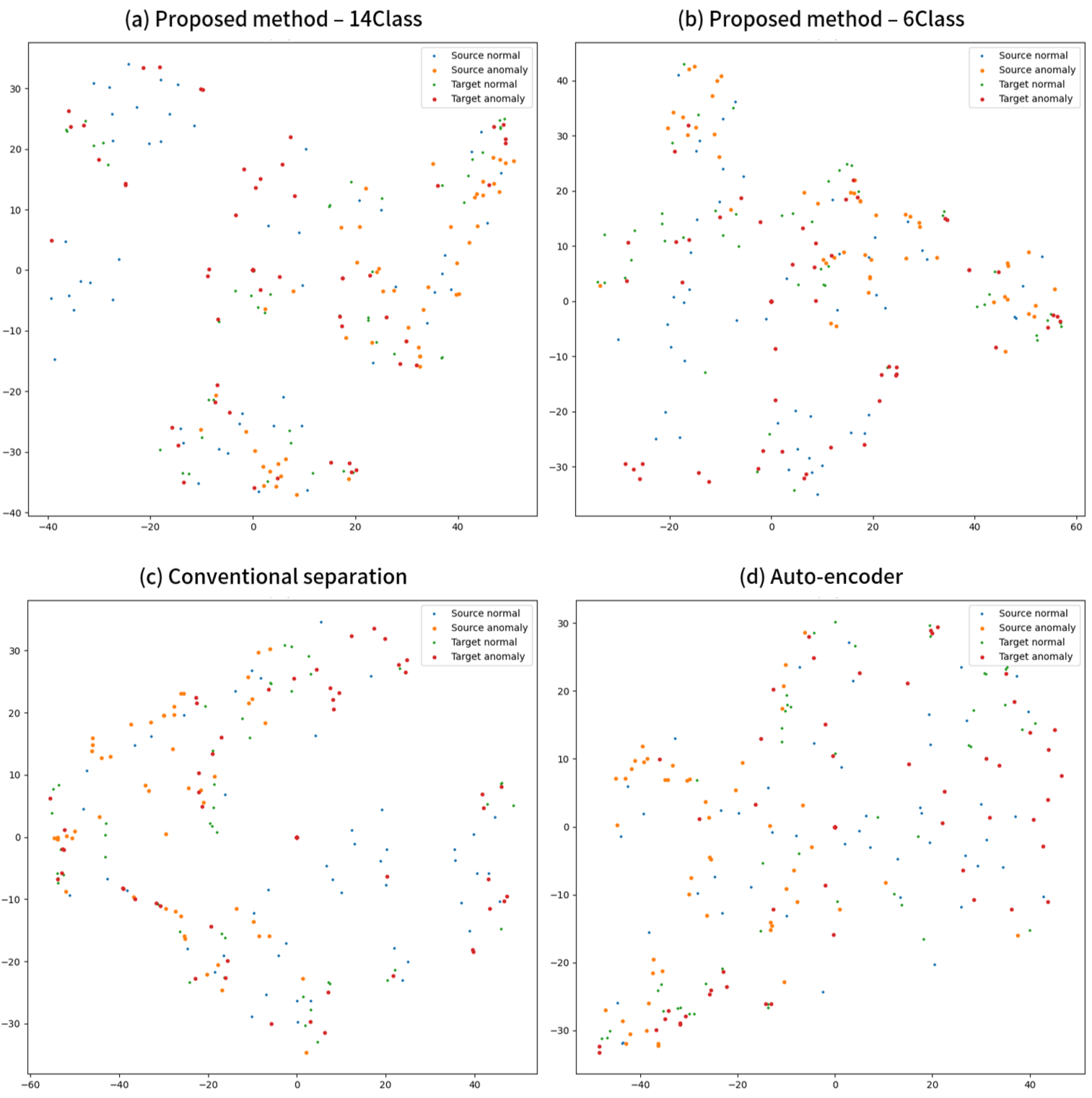}
    \caption{tSNE projection of representation of neural network trained by (a) Proposed method with 14 non-target classes (b) Proposed method with 6 non-target classes (c) Source separation trained to separate target class (d) Auto-encoder structure}
    \label{fig:tsne}
\end{figure*}

\section{Conclusions}
\label{sec:result}

This study proposes a novel approach to representation learning for anomalous detection systems, utilizing a neural network with a source separation model. Given the constraints of having only normal condition samples and limited training data, we developed a representation learning strategy that separates non-target class signals from a mixture of target and non-target class signals. Our method effectively leverages both the available training samples and data from other classes.
To evaluate the neural network's ability to learn representations that distinguish anomalous characteristics, we implemented an anomalous detection system that scores the obtained representations using the Mahalanobis distance and a maximum-likelihood covariance estimator. We compared our proposed method with alternative training strategies, including separating target signals from mixed sounds and estimating target signals from target-only inputs.
Results demonstrate that our method achieves superior performance, with a harmonic mean score of 54.58\%, compared to 53.99\% and 51.41\% for the alternative approaches. Notably, we observed that our training strategy yields improved representations with an increase in non-target class signals, even when the quantity of target class signals remains constant. Specifically, utilizing 14 non-target classes resulted in a score of 56.00\%, a 1.42\% improvement over the 6-class non-target scenario, and better results compared to the baseline methods (55.35\% and 55.02\%) employing two different anomalous scoring techniques.
Visualization of the learned representations using a t-SNE projection further corroborates the efficacy of our approach, revealing a more distinct separation between normal and anomalous samples compared to other methods.
In conclusion, we have proposed and validated a training strategy that effectively utilizes both target and non-target class samples. Our method of training neural networks to separate non-target signals from mixed inputs demonstrates better performance in obtaining target class representations compared to target separation and auto-encoder methods. Furthermore, we have shown that our approach can achieve enhanced representations with an increased diversity of non-target class signals, highlighting its potential for scalability and improved performance in anomalous sound detection tasks.

\bibliographystyle{IEEEtran}
\bibliography{refs}

%
%
%
%
%
%
%
%
%

\end{sloppy}
\end{document}